\documentstyle[sprocl]{article}

\def\Journal#1#2#3#4{{#1} {\bf #2}, #3 (#4)}

\def\NPA{{\em Nucl. Phys.} A}
\def\NPB{{\em Nucl. Phys.} B}
\def\PLB{{\em Phys. Lett.}  B}
\def\PRL{\em Phys. Rev. Lett.}
\def\PRC{{\em Phys. Rev.} C}
\def\PRD{{\em Phys. Rev.} D}
\def\ZPA{{\em Z. Phys.} A}

\def\be{\begin{equation}}
\def\ee{\end{equation}}
\def\bea{\begin{eqnarray}}
\def\eea{\end{eqnarray}}
\def\lsim{\mathrel{\rlap{\lower4pt\hbox{\hskip1pt$\sim$}}
    \raise1pt\hbox{$<$}}}         %less than or approx. symbol
\def\gsim{\mathrel{\rlap{\lower4pt\hbox{\hskip1pt$\sim$}}
    \raise1pt\hbox{$>$}}}         %greater than or approx. symbol

\begin{document}

\title{COHERENT NUCLEAR DIFFRACTIVE PRODUCTION OF MINI-JETS --- 
SIGNATURE OF COLOR TRANSPARENCY}

\author{G. A. MILLER}

\address{Department of Physics, Box 351560\\
University of Washington, Seattle, Washington 98195-1560, 
USA\\E-mail: miller@phys.washington.edu}

\maketitle\abstracts{The 
process $\pi + A$ leading to a pair of mini-jets at high relative 
transverse momentum ($k_t \gsim 2$ GeV) while leaving the nucleus in its 
ground state was selected as a definitive test of the existence of color 
transparency by Frankfurt, Miller and Strikman in 1993. The preliminary results 
of Fermilab experiment E791, led by Ashery and Weiss--Babai, show a strong 
A-dependence, consistent with the notion of 
color transparency.}

\section{Introduction}
The process we discuss is $\pi +A$ going to a $q \bar{q}$ pair of high relative 
transverse momentum ($k_t \gsim 2$ GeV) and leaving the nucleus in its ground 
state. The $q$ and $\bar{q}$ each decay into a mini-jet; I shall not discuss 
this aspect much. Observing the nuclear (A) dependence of this process was 
pointed out as a good way to study color transparency by us \cite{lfgam}
 in 1993. This 
was a long time ago. The reason for discussing this topic now is the recent 
apparent confirmation of the color transparency idea in Fermilab E791 by Ashery 
and Weiss--Babai \cite{wb}. 
The use of the very high energy $E_\pi = 500$ GeV beam 
makes these observations possible -- a small-sized wave packet can travel 
through the nucleus without expanding \cite{lfgamms}.

\section{The Process}
Why should color transparency occur in $\pi$ + A $\rightarrow q \bar{q} +$ A 
(ground state)? The selection of the    
final state as a $q \bar{q}$ pair moving at high relative momentum 
($k_t \gsim 2$ GeV)  plus the nuclear ground state causes the $q \bar{q}$ 
component of the pion to be the dominant component \cite{pion}. 
At 500 GeV, the pion 
breaks up into a $q \bar{q}$ pair well before hitting the target. Since $k_t$ 
is large, the transverse separation $b$ between the $q$ and $\bar{q}$ is small. 
The gluon emission from a color singlet $q \bar{q}$ pair (or any 
color--singlet, quark--gluon system) of small
spatial extent is suppressed\cite{gfb}. 
The forward scattering amplitude $f \sim i b^2 \sim i \sigma_0 
\frac{b^2}{<b^2>}$, where $\sigma_0$ is a normal size cross section ($\sim 
20-30$ mb) and $<b^2>$ is the pionic average of the squared transverse of the 
$q \bar{q}$ pair. The origin of $f$ is the exchange of two dipole gluons with 
the nuclear target. Two gluons are needed to maintain the color neutrality of 
the projectile and target. For large values of 
$k_t, f \sim -i \sigma_0 / ( <b^2> k^2_t )$. The $q \bar{q}$ 
production process (which simultaneously leaves the 
nucleon in its ground state) occurs rarely. Thus the dominant nuclear
amplitude 
occurs via only one interaction of $f$. There are no other initial or final 
state interactions because of the small size of the $q \bar{q}$ pair. This is 
color transparency. The single interaction of $f$ can occur on any nucleon 
and, because the final state is the nuclear ground state, is a coherent 
process. The nuclear amplitude ${\cal M}_A \propto Af$. This is a huge nuclear 
enhancement.

In the following, I'll discuss our 1993 calculation, the experiment and their 
relationship.

\section{Nucleon Amplitude}
The amplitude, ${\cal M} (N)$, for the process 
$\pi N\rightarrow q \bar{q} N$ 
 to occur on a nucleon $N$ is given, for large enough values of $k_t$ by
\begin{equation}
{\cal M} (N) = \int d^2 b\;\psi_\pi (x,b)\; {i \sigma_0}{b^2\over<b^2>} 
e^{-i \vec{k}_t \cdot \vec{b}} \; ,
\end{equation}
where $\psi_\pi (x,b)$ is the pion wave function and $x$ is the 
fraction of the pion longitudinal momentum carried by the final state quark. 
The anti-quark carries a fraction $1-x$. The $b^2$ operator can be replaced by 
$ -\nabla_{k_{t}}^2$ acting on the Fourier transform, 
$\tilde{\psi}_\pi (x, k_t)$, of the pion wave function. For wave functions 
with a power fall--off in $k_t$, the term $b^2$ then acts as $\sim - 1/k^2_t$. 
Note that $<b^2> \approx 0.24$ fm$^2$, while $1/k^2_t \sim 0.01 $fm$^2$, 
so the interaction is weak indeed. 
For the asympotic pion wave function we find \cite{lfgam}
\begin{equation}
{\cal M} (N) \sim \frac{i}{k^2_t} \frac{x (1-x)}{k^2_t} \alpha_s (k^2_t) \; .
\end{equation}

In this case
${\cal M} (N) \propto \tilde{\psi}_\pi (x,k_t)=\alpha_s (k^2_t )
\frac{x (1-x)}{k^2_t}$. 
This especially simple relation
occurs only if the final $q \bar{q}$ pair do not interact 
with each other by the exchange of a high momentum gluon \cite{mn}.

\section{Nuclear Amplitude}

The picture we have is that the pion becomes a $q \bar{q}$ pair of 
essentially zero transverse extent well before hitting the nuclear target. 
This point like configuration PLC can move through the entire nucleus without 
expanding. The $q \bar{q}$ can interact with one nucleon and can pass 
undisturbed through any other nucleon. For zero momentum transfer, $q_t$, 
to the nucleus, the amplitude ${\cal M}(A)$ takes the form 

\begin{equation}
{\cal M} {\rm (A)} = {\rm A} {\cal M} (N)
\left(1 + \frac{\epsilon}{<b^2> k^2_t} {\rm A}^{1/3}\right) 
\equiv {\rm A} {\cal M} (N) \gamma \; , 
\end{equation}

\noindent where the real number $\epsilon > 0$ and $\gamma > 1$. Observe the 
factor A which is the dominant effect here. The $\epsilon$ correction term 
arises from a soft rescattering which can occur as the PLC moves through the 
nuclear length $(R_A \propto$ A$^{1/3}$). The action of $f$ produces the 
$1/(<b^2> k^2_t)$ factor.

The differential cross section takes the form

\begin{equation}
\frac{d\sigma(A)}{dq^2_t} = A^2\gamma^2 \frac{d\sigma}{dq^2_\perp} 
(N) e^{-q^2_t R^2_A /3} \; .
\end{equation}
One measures \cite{wb} the integral
\begin{equation}
\sigma (A) = \int dq^2_\perp \frac{d\sigma (A)}{dq^2_t} = 
\frac{3}{R^2_A} A^2 \gamma^2 \sigma (N) \; .
\end{equation}
A typical procedure is to parametrize $\sigma(A)$ as 
\begin{equation}
\sigma (A) = \sigma_1 A^\alpha
\end{equation}
 in which $\sigma_1$ is a constant independent of A. For the 
$R_A$ corresponding to the two targets Pt (A = 195) and C(A = 12) of E791, 
one finds $\alpha \approx 1.45$, if $\gamma$ is taken to be unity. Including 
the value of $\gamma$ using the results of \cite{lfgam} leads to

\begin{equation}
\alpha \approx 1.58 \; .
\end{equation}

That $\gamma > 1$ was a somewhat surprising feature of our 1993 calculation 
because the usual second order scattering reduces cross sections. The key 
features of the usual first order term are: $f = i \sigma_0$ , and those of 
the second order term are $if^2 = -i\sigma^2_0$. Note the opposite signs. For 
us here $f = i \sigma_0 b^2/<b^2>$, which for very large values of $k^2_t$ 
becomes $f = -i \sigma_0 /( <b^2> k^2_t)$. The second order term depends on 
$if^2 = i [-i\sigma_0  / (<b^2> k^2_t)^2]^2$, which now has the same sign
as the first--order term.

\section{Requirements for Color Transparency}
These were discussed in Ref.~\cite{lfgam}. 
The two jets should have total transverse 
momentum to be very small. The relative transverse momentum should be 
$\gsim$ 2 GeV/c. One must identify the final nucleus as the target ground 
state. This is done \cite{wb} 
by isolating the $q^2_t$-dependence of the elastic 
diffractive peak. We need substantial A-dependence, $\sigma \equiv \sigma_1 
A^{1.6}$, for large enough values of $k_t$. The background processes involving 
nuclear excitation vary as A, so an unwanted counting of such would actually 
weaken the signal we seek. The amplitude varies as 
${\cal M}(A) \sim \alpha_s / k^4_t$ and $\sigma (A) \sim \alpha^2_s / k^8_t$ ; 
this should be checked experimentally. For the amplitude discussed here, 
$\sigma (A) \sim ( x(1-x))^2$. See however \cite{mn}.

\section{The experiment -- ELAB E791}
All of the information I have is from Ashery's webpage\cite{wb}. 
This discusses the
excellent resolution of the transverse momentum,  shows the 
identification of the di-jet using the Jade algorithm, and displays
the identification 
of the diffractive peak by the $q_t^2$ dependence for very low $q^2_t$. This 
dependence is consistent with that obtained from the previously measured radii 
$R_c = 2.44$ fm, and $R_{Pt} = 5.27$ fm.

Their preliminary result is
\begin{equation}
\alpha \approx 1.55 \pm 0.05 \; ,
\end{equation}
 which is remarkably close to the theoretical value shown in Eq.(9). 
The $k^2_t$ dependence of the scattering amplitude has not been checked, but 
we are told that this is both feasible and in progress.

The measured cross section does have an $x$-dependence which is reasonably well 
described by $[x(1-x)]^2$. This lends support to the notion that the 
asymptotic behavior is manifest in nature. However, ${\cal M} (A)$ is only 
approximately proportional to the quark distribution amplitude \cite{mn}.

\section{Electromagnetic Background}

Because very low values of $q^2_t$ are involved, one could ask if the process 
occurs by one photon exchange (Primakoff effect) instead of two gluon exhange. 
Even if the Primakoff process were dominant, the unusual A-dependence would be 
a consequence of color transparency. However, we can estimate the relative 
importance of the two effects. We have

\begin{equation}
{\cal M}_P (N) = \frac{e^2 \tilde{\psi}_\pi (x,k_t) Z}{q^2_t} \; ,
\end{equation}
 which should be compared with the amplitude of Eq.~(2) written as
\begin{equation}
{\cal M} (N) = \tilde{\psi}_\pi (x,k_t) A \frac{\sigma_0}{<b^2>} 
\frac{i}{k^2_t}
\end{equation}
 using $q^2_t \approx 0.02$ GeV$^2,$  Z = 78, $e^2 = 1/137 , 
k_t = 2$ GeV, and\cite{lfgam}$\sigma_0 / <b^2> \approx 10$  gives
\begin{equation}
\frac{{\cal M}_P (N)}{{\cal M} (N)} = -0.06 i \; .
\end{equation}
The Primakoff term is small and, because of its real nature, does not 
interfere with the larger stong amplitude.

\section{Meaning of $\alpha$ = 1.55}

The use of $\alpha = 1.55$ in Eq.~(6) leads to
\begin{equation}
\frac{\sigma (Pt)}{\sigma (C)} = 75 \; .
\end{equation}
The usual dependence of a diffractive process is $\approx A^{2/3}$ or
$ \alpha = 2/3$. This would give
\begin{equation}
\frac{\sigma_{\rm USUAL} (Pt)}{\sigma_{\rm USUAL} (C)} = 7.
\end{equation}
Thus color transparency causes a factor of 10 enhancement! This seems to be 
the huge effect of color transparency that many of us have been hoping to 
find. A word of caution must be sounded. The $k^2_t$ dependence of the cross 
section must be verified at least roughly to be consistent with the 
$(\alpha_s / k^4_t)^2$ behavior discussed above.

\section{Implications}  
Suppose color transparency has been correctly observed in 
$\pi + A \rightarrow q \bar{q} + A$ (ground state). So what?

There are many implications.
The spectacular enhancement of the cross section would be  a new 
novel effect. The 
point like configurations PLC  would be proved to exist.
This is one more verification of the concept and implications of the idea
of color.
 Previous experiments \cite{asc} showing hints \cite{bkj} 
of color transparency (for a review
 see Ref~\cite{lfgamms}) probably do 
show color transparency. Efforts \cite{ke}  
to observe color transparency at Jefferson
Lab should be increased.

The idea that a nucleon is a composite object is emphasized by these findings. 
If there are PLC, there must also exist blob--like configurations of Huskyons 
\cite{ab}. These 
different configurations have wildly different interactions with a 
nucleus \cite{llf}, so 
that the nucleon in the nucleus can be very different from a 
free nucleon. This leads to an entirely new view of the nucleus, one in which 
the nucleus is made out of oscillating, pulsating, vibrating, color singlet, 
composite objects.
\section*{Acknowledgments}
This work is partially supported by the USDOE, and is
based on a collaboration with L. Frankfurt and M. Strikman.

\end{document}